# Secrecy Capacity and Energy Efficiency of Spectrum Sharing Networks Incorporating Physical Layer Security

Yee-Loo Foo

*Faculty of Engineering, Multimedia University*
*Persiaran Multimedia, 63100 Cyberjaya, Selangor, Malaysia*
ylfoo@mmu.edu.my
+603-8312-5376
ORCiD 0000-0001-6106-4655

*Abstract*—Underutilized wireless channel is a waste of spectral resource. Eavesdropping compromises data secrecy. How to overcome the two problems with one solution? In this paper, we propose a spectrum sharing model that defends against eavesdropping. Consider a source-destination channel that is being eavesdropped. Cognitive radio can help jamming the eavesdropper. Friendly jamming is a physical layer security method of protecting data secrecy based on radio propagation characteristic. In return, the helper is allowed to access the source's unused channel. The desirable characteristic of cognitive radio is its capability of sensing the occupancy/vacancy of the channel. This work investigates the secrecy capacity $C_S$ and energy efficiency $\mu$ of such a spectrum sharing network that deploys physical layer security method. The main factors that affect $C_S$ and $\mu$ are the transmit powers of the source and cognitive radio. We present a novel expression that permits finding the sensing duration $t$ that optimizes $\mu$.

## I. INTRODUCTION

When a wireless transmitter is idle, the channel resource assigned to it becomes waste. When transmission begins, the channel may be eavesdropped. We propose in this paper a spectrum sharing model that alleviates the problems of spectral wastage and eavesdropping. It makes use of a spectrum sensing cognitive radio. When an idle channel is found through the sensing process, the cognitive radio acquires the channel for its own transmission. On the other hand, when the channel's incumbent user becomes active again and starts transmission, there is a payback that the cognitive radio must transmit artificial noise to jam the eavesdropper. This is called friendly jamming, a concept of physical layer security [1][2][3]. Unlike the cryptographic security methods, physical layer security measure that makes use of the radio propagation characteristic is simpler and it does not required a secret key [3][4]. In fact, physical layer security method is widely regarded as a solution to the security of the next generation wireless networks [1][2][3]. This paper utilizes both the ideas of spectrum sharing and physical layer security.

In the context of spectrum sharing, the incumbent user is the primary user (PU), the cognitive radio is the secondary user (SU). Conventionally PU does not like sharing the spectrum, because SU's transmission may cause interference. What we propose here is to give an incentive to the PU, that the SU will help defending against eavesdropping when a primary transmitter (PT) sends data to a primary receiver (PR). A cognitive radio transmitter (CT) transmits artificial noise that is decodable by only the PR but not an eavesdropper (EA). In return, CT receives the benefit of accessing the channel for sending its own data to a cognitive radio receiver (CR) when PT stops transmitting. This scenario is beneficial to both the PU pair (PT and PR) and SU pair (CT and CR). Fig. 1 illustrates the nodes involved and the distances between them.

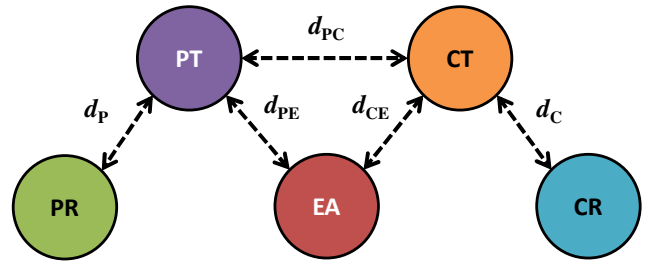

Fig. 1. The nodes and the distances between them

From the perspective of physical layer security, PT is the *source* of data to be protected, whereas PR is the *destination*. CT is employed as a friendly jammer or *helper*. The adversary *eavesdropper* (EA) targets a selected victim. It is reasonable to assume that the *eavesdropper* itself has constraints of energy and processing power. It cannot eavesdrop unselectively all nodes in the surroundings. The *source* (PT) is the victim of eavesdropping. The *source* must engage an untargeted node to be the *helper* (in this case, CT). Deployment of a helper can improve the secrecy capacity $C_S$, which is defined as the maximum rate of sending out a secure message without being decoded by the EA. CT transits between three states: sensing, jamming the EA (when the channel is occupied), and transmitting CT's data (when the channel is relinquished by PT). CT allocates its resources (time and energy) between three states: sensing, jamming, and transmitting data. In a frame of 1 second, suppose CT spends $t$ fraction of it for sensing, the other fraction 1-$t$ for either jamming the EA or transmitting its own data.

Our proposed model encourages the incumbent to share the spectrum since PR receives protection from the CT. *But what benefit does the SU pair gain*? SU's throughput gains because of the gain in spectral opportunity. Nevertheless, CT may consume excessive energy in jamming the EA. This is risky especially if CT has energy constraint (e.g. battery-powered). *How to achieve high throughput while consuming minimal amount of energy*? It is important to answer this question, since it is critical for CT to remain highly energy efficient. Energy efficiency $\mu$ is the ratio of achievable throughput to the total energy consumed [5][6]. CT could adjust its transmit power ($P_C$) and sensing duration ($t$) in order to attain an optimal $\mu$. For example, enlarging $t$ increases the sensing accuracy, allowing CT to make a more accurate judgement about the transmission status of PT. Ideally such action will increase the throughput and reduce the energy consumption, resulting in higher $\mu$. Nevertheless, enlarging $t$ also reduces the time portion allocated for transmitting data and consequently the throughput, implying a reduced $\mu$. These conflicting tendencies actually result in an optimal $\mu$, and our study here proves that such optimum exists.

This paper investigates the secrecy capacity $C_S$ and energy efficiency $\mu$ of a spectrum sharing network. We have reached a few important conclusions in the end. The required $P_C$ increases with the target secrecy rate $R_S$. Nevertheless, if PT transmits at strong $P_P$ level, much smaller $P_C$ is needed. On the other hand, $t$ needs to be carefully chosen in order to maximize $\mu$. Let $t^*$ denote the $t$ value that results in the optimal energy efficiency, $\mu^*$. We present a mathematical expression that permits finding the $\{t^*, \mu^*\}$ pair. At the end of our analyses, we find that both $\mu^*$ and $C_S$ increase with $P_P$. $C_S$ increases with $P_C$ too, but this is not the case for $\mu^*$. Although $\mu^*$ increases with $P_C$ for $P_C < 0.5$ W, it decreases exponentially with $P_C$ for larger $P_C$ values.

The remainder of the paper is organized as follows. Section 2 gives a thorough literature review in the areas of spectrum sharing, energy efficiency, and physical layer security. Section 3 analyzes the requirement of achieving satisfactory secrecy capacity, whereas Section 4 analyzes the requirements of attaining the maximal energy efficiency. In the end, Section 5 draws the conclusion.

## II. LITERATURE REVIEW

### A. Spectrum Sharing

At the turn of the millennium, spectrum scarcity has become an issue and a heated topic for discussion. Peha [7] summarized the two mainstream of views on possible solutions, i.e. to treat the spectrum as a property or something that can be shared. Cabric at el. [8] investigated different radio designs for the purpose of spectrum sharing. (1) Ultra Wideband (UWB) radio that transmits extremely low power (that it does not interfere with any existing primary transmission) over a very large bandwidth - the *Underlay* concept. (2) Cognitive Radio with spectrum sensing capability that can dynamically avoid any channels that are actively accessed by the primary users (PUs) - the *Overlay* concept. Ghasemi and Sousa [9] suggested the cognitive radio to leverage on deep fading effect to opportunistically access the shared channel. When the cognitive radio's signal is severely faded at the primary receiver, fading becomes an advantage. They further investigated the relationship between the cognitive radio's channel capacity and the level of interference experienced at the primary receiver. In a multiuser system, those causing severely faded interfering signals can be selected to transmit, resulting in a multiuser diversity gain [10]. Ban et al. [10] have analyzed and characterized such diversity gain.

There are various methods of sharing the spectrum. Cao and Zheng [11] proposed a local bargaining method for assigning spectrum to mobile users of an ad-hoc network of which the topology may often change. Huang et al. [12] suggested auctioning that the user bid for a desired portion of spectrum. The bidding constitutes a game. Etkin et al. [13] aimed to achieve spectral efficiency and fairness through a the use of a distributed set of self-enforced rules. The users play a repeated game that allows them to build reputation or be punished for inappropriate behavior. Etkin et al. [13] demonstrated the possibility of achieving an efficient and fair spectrum sharing goal. Ji and Liu [14] proposed using a belief-assisted pricing approach to allocate spectrum efficiently to the secondary users (SUs) while countering the selfish behavior of the PUs. All the aforementioned games models had been analyzed and compared [15]. The players of a spectrum allocation game can be the primary instead of the SUs [16]. In this scenario, the PUs try to maximize their profits by 'selling' available spectra to the SUs at competitive prices. The cost of doing so is the degradation of the quality of primary service.

Keeping the transmission powers of the SUs under control can essentially lower the level of interference experienced by the PUs. Kang et al. [17] proposed adjusting the transmission power following the result of spectrum sensing. They investigated the achievable capacity of secondary channel with respect to the changes in transmission power and sensing duration. They proved the existence of an optimal power allocation and sensing duration.

Multi-antennas can be deployed to enhance the capacity of a secondary channel due to the spatial gain and by directionally avoiding the PUs [18]. Zhang and Liang [18] have characterized the secondary channel capacity and provided the optimal designs for beamforming and spatial multiplexing under different power and interference constraints. Xie et al. [19] have considered an even more complex system that may consist of hundreds of antennas, i.e. massive multiple-input, multiple-output (MIMO) system. These antennas can beam at different azimuth angles and elevations, resulting in a three dimensional MIMO system.

There are certainly many different ways of sharing the spectrum. None of the aforementioned works concern with protecting the secrecy of data transmission. One of the related works that is closest to ours is [20]. Zou [20] investigated physical layer security in a spectrum-sharing network, where none of the users is given the priority of channel access. This is different from the setting here where we prioritize PU over SU, with the intention of investigating the energy efficiency of SU.

### B. Energy Efficiency

The energy efficiency of SU has always been subject to an interference constraint imposed by PU. In proposing a spectrum sharing model that encourages a PU share the spectrum, we need to ensure that it is not disadvantageous to the SU. One of the main concerns is the SU's energy efficiency, which is a big issue in cognitive radio networks (CRNs) [5][21][22][23][24][25]. However, energy efficiency has not been considered in the context of SU being required to protect PU's secrecy. In order to achieve the highest energy efficiency, Pei et al. [5] optimized the channel sensing order, timing of sensing and transmitting, and transmission power of SU. Deng et al. [21] optimized the scheduling of spectrum sensing with the aim of minimizing energy consumption. Xie et al. [22] proposed energy-efficient spectrum and power allocations between macrocells and femtocells of a heterogeneous CRN architecture, which is different from ours. The CRN structure considered in [23] is also different. It consisted of a cognitive base station that allocated frequency resources to the SUs. Optimal scheduling of such allocation was proposed in order to maximize energy efficiency. We certainly welcome the idea of 'greening' the CRN by capturing and storing energy from the ambience [24]. Nevertheless, this work focuses on saving energy rather than harnessing energy. Costa and Ephremides [25] evaluated how energy efficiency tradeoffs with throughput and sensing accuracy. However, the schemes they have considered are not entirely similar to ours. Two major spectrum sharing schemes have been considered in [25]: (1) Underlay – SUs transmit simultaneously as PU does, subject to certain power constraint. (2) Interweave – If the channel is sensed to be idle, a SU transmits with a finite probability. The third scheme is the hybrid of both. All the aforementioned works are concerned with the energy efficiency of SU subject to an interference constraint imposed by PU, but not security constraint.

### C. Physical Layer Security

Physical layer security is a method that protects data secrecy. Although sharing the same objective as cryptography, the method is different that it exploits the noisy characteristic of a channel. Briefly, if an eavesdropper's (wiretap) channel is noisier than the legitimate channel, then the eavesdropper has no advantage in deciphering any messages passed around between two legitimate users. The theory of physical layer security is founded on the seminal work by Wyner [4] on a degraded broadcast channel. Leung and Hellman [26] had extended the concept to a Gaussian channel. The same concept is further applied to wireless (fading) channels [27]. Being defined as the maximum rate at which a sender can transmit to a receiver data that is not intercepted by an eavesdropper, the secrecy capacity varies subject to the fading condition.

Spectrum resources are wasted when they are unused by the licensed or primary users (PU). Cognitive radio network (CRN) could opportunistically access the unused spectrum bands, thus improving spectrum utilization. The unlicensed cognitive radio users are to perform spectrum sensing in order to identify unused bands [28]. However, CRN is expected to contribute in enhancing PU's performance, and then it is rewarded with spectral opportunities in return [28]. Such a cooperative cognitive radio network could help the PU in many ways.

Conventionally, cryptographic methods are used to protect data secrecy. However, the legitimate data is still exposed to the eavesdropper who may record it to be decoded later when it obtains higher computation power. Physical layer security method resolves the security issue at the physical layer itself, by not letting the eavesdropper having the message at the first place. To tackle eavesdropping, a friendly jammer can transmit artificial noise (AN) to jam the eavesdropper's reception. The idea of cooperatively jamming the eavesdropper by using artificial noise (AN) was introduced in [29][30][31]. In [29], a multi-antenna transmitter or its amplify-and-forward (AF) relays were eavesdropped. The transmitter allocated a portion of its power for generating artificial noise in order to degrade the eavesdropping channel. The artificial noise was selected from the null space of the legitimate receiver's channel but not that of the eavesdropper's channel. It thus caused adverse effect to the eavesdropper's reception but not the legitimate receiver's.

The idea of transmitting artificial noise to jam the eavesdropper is further extended in [32][33][34]. Subject to a PU secrecy constraint, Xu and Li [32] maximized the CT ergodic transmission rate through optimal scheduling and power allocation among CTs. Further in [33], they maximized CT ergodic secrecy rate with the CTs were also being eavesdropped. In both [32][33], the unscheduled CTs remained idle. In [34], the unscheduled CTs sent artificial noise to frustrate the eavesdroppers, with the objective of minimizing PT secrecy outage. Our objective here is to maximize the energy efficiency of CT.

A thorough literature survey in the three aforementioned areas reveals the need to investigate the energy efficiency and secrecy capacity of a spectrum sharing network that deploys physical layer security technique to counter eavesdropping.

## III. ANALYSIS OF SECRECY CAPACITY

We have described the spectrum sharing model in Section 1. In this section, we analyze the secrecy capacity, $C_S$. Let $\gamma_{PC}$ denote the signal to noise power ratio (SNR) measured at the CT due to **PT**'s signal. Suppose **CT** consumes $P_C$ watts and **PT** transmits at the power of $P_P$ watts within the 1-second frame. The SNR at the CT caused by PT's signal, $\gamma_{PC}$ is as follows:

$$\gamma_{PC} = \frac{|h^2|}{N} \cdot \frac{P_P}{d_{PC}^\alpha} \qquad (1)$$

$N$ denotes the noise power measured at the CT, $h$ the channel gain, $d_{PC}$ the PT-CT distance, and $\alpha$ the path loss exponent. In similar manner, the SNR at the CR due to CT's data transmission, $\gamma_C$ is given by

$$\gamma_C = \frac{|h^2|}{N} \cdot \frac{P_C}{d_C^\alpha} \qquad (2)$$

Here we assume that $N$ and $h$ are the same, $d_C$ denotes the CT-CR distance. Similarly, the SNR at the PR caused by PT's data transmission, $\gamma_P$ is as follows:

$$\gamma_P = \frac{|h^2|}{N} \cdot \frac{P_P}{d_P^\alpha} \qquad (3)$$

We assume that $N$ and $h$ are the same for PR, $d_P$ denotes the PT-PR distance. Let $\gamma_{PE}$ denote the SNR measured at the **EA** caused by **PT**'s data transmission, $\gamma_{CE}$ the SNR at **EA** caused by **CT**'s artificial noise, $d_{PE}$ the PT-EA distance, and $d_{CE}$ the CT-EA distance.

$$\gamma_{PE} = \frac{|h^2|}{N} \cdot \frac{P_P}{d_{PE}^\alpha} \qquad (4)$$

$$\gamma_{CE} = \frac{|h^2|}{N} \cdot \frac{P_C}{d_{CE}^\alpha} \qquad (5)$$

We make similar assumption regarding $N$ and $h$.

Let $C_P$ denote the channel capacity normalized by bandwidth of the primary channel (i.e. PT-PR link), $C_C$ the normalized capacity of the cognitive radio channel (i.e. CT-CR link), and $C_E$ the normalized capacity of the eavesdropper channel (i.e. PT-EA link).

$$\begin{aligned}C_P &= \log_2(1+\gamma_P) \\ &= \log_2\left(1+\frac{|h^2|}{N}\cdot\frac{P_P}{d_P^\alpha}\right)\end{aligned} \qquad (6)$$

$$\begin{aligned}C_C &= \log_2(1+\gamma_C) \\ &= \log_2\left(1+\frac{|h^2|}{N}\cdot\frac{P_C}{d_C^\alpha}\right)\end{aligned} \qquad (7)$$

$$\begin{aligned}C_E &= \log_2\left(1+\frac{\gamma_{PE}}{1+\gamma_{CE}}\right) \\ &= \log_2\left(1+\frac{\frac{P_P}{d_{PE}^\alpha}}{\frac{N}{|h^2|}+\frac{P_C}{d_{CE}^\alpha}}\right)\end{aligned} \qquad (8)$$

**Secrecy capacity**, $C_S$ is the maximum rate (normalized by bandwidth) of PT sending a secure message to PR without being decoded by EA.

$$\begin{aligned}C_S &= C_P - C_E \\ &= \log_2(1+\gamma_P) - \log_2\left(1+\frac{\gamma_{PE}}{1+\gamma_{CE}}\right) \\ &= \log_2\left(\frac{1+\gamma_P}{1+\frac{\gamma_{PE}}{1+\gamma_{CE}}}\right)\end{aligned} \qquad (9)$$

$C_S > R_S$, where $R_S$ represents the target secrecy rate. $R_S > 0$ for secure transmission. Otherwise, data secrecy is compromised.

$$C_{S,\,min} = R_S$$

Given certain $R_S$, we can relate the minimum required $P_C$ in response to the changes in $P_P$.

$$\frac{1+\gamma_P}{1+\frac{\gamma_{PE}}{1+\gamma_{CE,\,min}}} = 2^{R_S} \qquad (10)$$

After some substitutions into (10), we find that

$$\frac{|h^2|}{N}\cdot\frac{P_P}{d_P^\alpha} - \frac{\frac{|h^2|}{N}\cdot\frac{2^{R_S} P_P}{d_{PE}^\alpha}}{1+\frac{|h^2|}{N}\cdot\frac{P_{C,\,min}}{d_{CE}^\alpha}} = 2^{R_S} - 1$$

Finally, we can express $P_{C,\,min}$ in terms of $P_P$ and other parameters as follows,

$$P_{C,\min} = \frac{a}{\dfrac{1}{d_P^\alpha} - \dfrac{b}{P_P}} \quad (11)$$

where $a = \dfrac{2^{R_S}\left(\dfrac{d_{CE}}{d_{PE}}\right)^\alpha}{1+\dfrac{|h^2|}{N}}$, $b = \dfrac{2^{R_S}-1}{\dfrac{|h^2|}{N}}$

Fig. 2 reveals the relationship between $P_{C,\min}$ and $P_P$, given different requirements of $R_S$. To plot Fig. 2, we assume that $\alpha = 2$, $\dfrac{|h^2|}{N} = 10$. In addition, we assume equidistance among the nodes, i.e.

$$d_C = d_P = d_{PC} = d_{PE} = d_{CE}$$

We arbitrarily assume that all the distances above are 2 m.

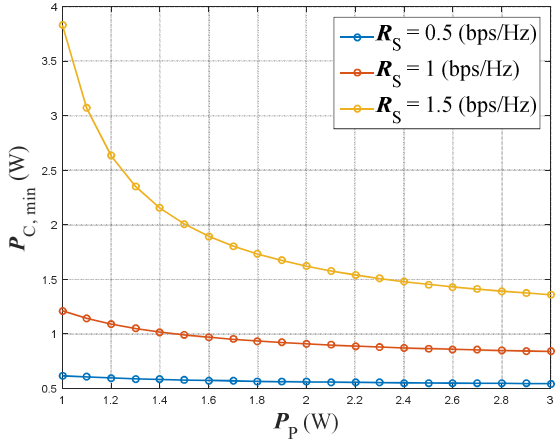

Fig. 2. $P_{C,\min}$ vs. $P_P$

Naturally, when $R_S$ becomes more demanding, $P_{C,\min}$ becomes higher. When $P_P$ is high enough, smaller $P_{C,\min}$ is required.

## IV. ANALYSIS OF ENERGY EFFICIENCY

We now analyze the spectrum sensing performance. We adopt energy detector for spectrum sensing due to its tractable analysis. Let us consider a simple case of a 1-second frame. CT spends $t$ fraction of the time for spectrum sensing. If the sampling rate is $f_s$, the number of collected energy samples is $f_s t$. The central limit theorem applies when there is a large number of energy samples [35]. Given a *target* probability of detection $\hat{\Pr}_D$, the probability of false alarm $\Pr_F$ can be written as

$$\Pr_F = Q\left(A + B\sqrt{t}\right) \quad (12)$$

where $A = Q^{-1}(\hat{\Pr}_D)\sqrt{1+2\gamma_{PC}}$, $B = \gamma_{PC}\sqrt{\dfrac{f_s}{2}}$

$Q(\cdot)$ function is the standard Gaussian complementary cumulative distribution function (CDF), whereas $Q^{-1}(\cdot)$ is its inverse.

In the remaining 1-$t$ fraction of the 1-second duration, CT transmits its data to CR if PT is sensed to be idle. However, if PT is found to be actively transmitting, CT enters into the counter-eavesdropping phase of generating artificial noise to jam EA. The total energy consumed in 1 second is $P_C$ joules. The sampling rate $f_s$ is directly proportional to $P_C$. Assume that

$$f_s = kP_C \text{ sample/s, } k \text{ is a constant.} \quad (13)$$

Let $\Pr_1$ and $\Pr_0$ denote the respective probabilities of PT actually transmitting and not transmitting, $\Pr_{(1)}$ and $\Pr_{(0)}$ the respective probabilities of CT perceiving PT to be transmitting and not transmitting. Note that perception may not represent reality. Two conditions contribute towards $\Pr_{(0)}$, i.e. (1) if the PT is not transmitting and there is no false alarm, and (2) if the PT is actively transmitting but there is a miss detection [6]. Thus $\Pr_{(0)}$ is given by

$$\Pr_{(0)} = \Pr_0(1-\Pr_F) + \Pr_1(1-\hat{\Pr}_D) \quad (14)$$

Substituting (12) into (14),

$$\Pr_{(0)} = 1 - \Pr_1\hat{\Pr}_D + (\Pr_1-1)Q\left(A+B\sqrt{t}\right) \quad (15)$$

where $A = Q^{-1}(\hat{\Pr}_D)\sqrt{1+2\gamma_{PC}}$, $B = \gamma_{PC}\sqrt{\dfrac{f_s}{2}}$

Likewise, $\Pr_{(1)}$ is contributed by two other conditions: (1) if the PT is not transmitting but there is a false alarm, and (2) if the PT is actively transmitting and there is no miss detection.

$$\Pr_{(1)} = 1 - \Pr_{(0)}$$

The probability of CT counter-eavesdropping is given by $\Pr_{(1)}$. When CT senses PT's transmission, CT refrains from transmitting data. Instead, CT generates artificial noise to jam EA in order to protect the secrecy of PT's data. The achievable throughput of the cognitive radio channel [35] is given by

$$R_C = (1-t)\Pr_{(0)} C_C \quad (16)$$

The first fraction of $t$ second during which CT performs sensing does not contribute to the throughput. Energy efficiency $\mu$ is the ratio of achievable throughput to the total energy consumed [5][6],

$$\mu = \frac{R_C}{P_C} = \frac{C_C}{P_C}(1-t)\Pr_{(0)} \quad (17)$$

Substituting (15) into (17),

$$\mu = \frac{C_C}{P_C}(1-t)\left[1-\Pr_I \hat{\Pr}_D + (\Pr_I -1)Q\left(A+B\sqrt{t}\right)\right] \quad (18)$$

where $A = Q^{-1}(\hat{\Pr}_D)\sqrt{1+2\gamma_{PC}}$, $B = \gamma_{PC}\sqrt{\dfrac{f_s}{2}}$

To find max $\mu$, $\dfrac{d\mu}{dt} = 0$

After some differentiations and substitutions, we find that

$$\left(\frac{1}{\sqrt{t}} - \sqrt{t}\right)\frac{(1-\Pr_I)B}{2\sqrt{2\pi}}e^{\frac{-D^2}{2}} - 1 \\ + \Pr_I \hat{\Pr}_D + (1-\Pr_I)Q(D) = 0 \quad (19)$$

where $D = A + B\sqrt{t}$, $A = Q^{-1}(\hat{\Pr}_D)\sqrt{1+2\gamma_{PC}}$, $B = \gamma_{PC}\sqrt{\dfrac{f_s}{2}}$

Using numerical method, we can find $t$ that solves (19). $t$ that satisfies (19), i.e. $t^*$ will result in maximum $\mu$, i.e. $\mu^*$.

Fig. 3 proves the existence of $t^*$, which is generated from the following values: $P_C = P_P = 1$ W, $\Pr_1 = 0.3$, $\hat{\Pr}_D = 0.9$, $\alpha = 2$, $\dfrac{|h^2|}{N} = 10$, and $k = 100$.

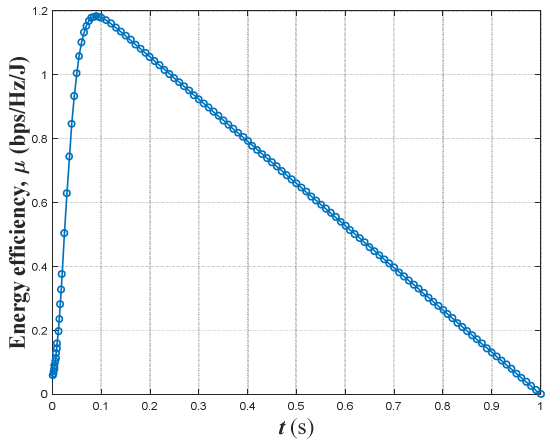

Fig. 3. $\mu$ vs. $t$

Fig. 3 clearly reveals that: there is a $t$ value that maximizes $\mu$. In order to have better understanding on the maximum energy efficiency $\mu^*$, we proceed to evaluate how $\mu^*$ varies with two important parameters, $P_C$ and $P_P$. These two powers affect both $\mu^*$ and $C_S$.

First, let us evaluate how $\mu^*$ and $C_S$ vary with $P_C$. Fig. 4 is generated from $P_P = 1$ W, $\Pr_1 = 0.3$, $\hat{\Pr}_D = 0.9$, $\alpha = 2$, $\dfrac{|h^2|}{N} = 10$, and $k = 100$. Naturally, the secrecy capacity $C_S$ increases with $P_C$, which benefits the primary users. However, such action reduces the achievable $\mu^*$ (for $P_C > 0.5$ W), which is the price to pay.

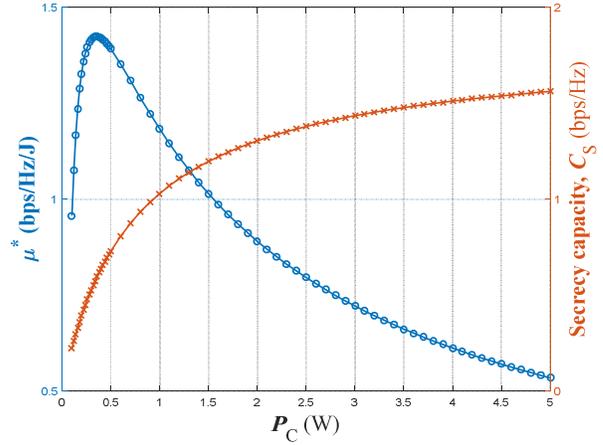

Fig. 4. $\mu^*$ and $C_S$ vs. $P_C$

Now let us evaluate how $\mu^*$ and $C_S$ vary with $P_P$. Fig. 5 is generated from $P_C = 1$ W, $\Pr_1 = 0.3$, $\hat{\Pr}_D = 0.9$, $\alpha = 2$, $\dfrac{|h^2|}{N} = 10$, and $k = 100$. By intuition, the secrecy capacity $C_S$ increases with $P_P$. $\mu^*$ increases with $P_P$ too, but only to certain extent. For $P_P > 1.5$ W, $\mu^*$ no longer increases with $P_P$. This is because the increment in $P_P$ improves the detection accuracy. After certain level of accuracy is attained, the energy efficiency depends more on other factors.

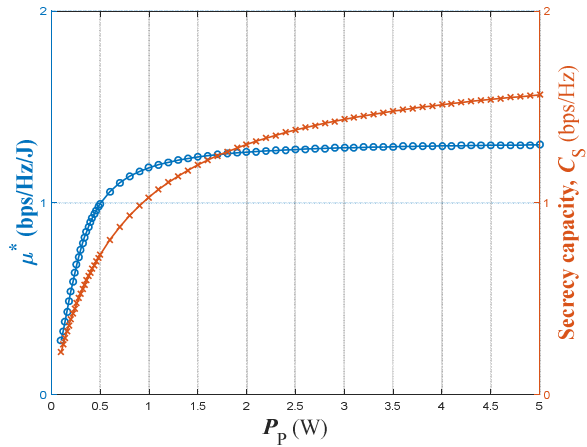

Fig. 5. $\mu^*$ and $C_S$ vs. $P_P$

## V. Conclusion

This paper investigates the secrecy capacity and energy efficiency of a spectrum sharing network. We have reached a few important conclusions. The required $P_C$ increases with the target secrecy rate $R_S$. Nevertheless, if PT transmits at strong $P_P$ level, much smaller $P_C$ is needed. On the other hand, $t$ needs to be carefully chosen in order to maximize $\mu$. We present eq. (19) that permits finding the $\{t^*, \mu^*\}$ pair. Both $\mu^*$ and $C_S$ increase with $P_P$. $C_S$ increases with $P_C$ too, but this is not the case for $\mu^*$. Although $\mu^*$ increases with $P_C$ for $P_C < 0.5$ W, it decreases exponentially with $P_C$ for larger $P_C$ values. For future works, we will analyze the requirements of countering an active jammer instead of eavesdropper.

**Conflict of Interest Statement**

*On behalf of all authors, the corresponding author states that there is no conflict of interest.*

## References

[1] A. Mukherjee, S. A. A. Fakoorian, J. Huang, and A. L. Swindlehurst, "Principles of physical layer security in multiuser wireless networks: A survey," *IEEE Commun. Surv. Tutorials*, vol. 16, no. 3, pp. 1550–1573, 2014, doi: 10.1109/SURV.2014.012314.00178.

[2] Y. Liu, H. H. Chen, and L. Wang, "Physical Layer Security for Next Generation Wireless Networks: Theories, Technologies, and Challenges," *IEEE Commun. Surv. Tutorials*, vol. 19, no. 1, pp. 347–376, Jan. 2017, doi: 10.1109/COMST.2016.2598968.

[3] H. V. Poor and R. F. Schaefer, "Wireless physical layer security," *Proc. Natl. Acad. Sci. U. S. A.*, vol. 114, no. 1, pp. 19–26, Jan. 2017, doi: 10.1073/pnas.1618130114.

[4] A. D. Wyner, "The Wire-Tap Channel," *Bell Syst. Tech. J.*, vol. 54, no. 8, pp. 1355–1387, 1975, doi: 10.1002/j.1538-7305.1975.tb02040.x.

[5] Y. Pei, Y. C. Liang, K. C. Teh, and K. H. Li, "Energy-efficient design of sequential channel sensing in cognitive radio networks: Optimal sensing strategy, power allocation, and sensing order," *IEEE J. Sel. Areas Commun.*, vol. 29, no. 8, pp. 1648–1659, Sep. 2011, doi: 10.1109/JSAC.2011.110914.

[6] S. Althunibat and F. Granelli, "On Results' Reporting of Cooperative Spectrum Sensing in Cognitive Radio Networks," *Telecommun. Syst.*, vol. 62, no. 3, pp. 569–580, Jul. 2016, doi: 10.1007/s11235-015-0095-5.

[7] J. M. Peha, "Approaches to spectrum sharing," *IEEE Commun. Mag.*, vol. 43, no. 2, Feb. 2005, doi: 10.1109/MCOM.2005.1391490.

[8] D. Cabric, I. D. O'Donnell, M. S. W. Chen, and R. W. Brodersen, "Spectrum sharing radios," *IEEE Circuits Syst. Mag.*, vol. 6, no. 2, pp. 30–45, 2006, doi: 10.1109/MCAS.2006.1648988.

[9] A. Ghasemi and E. S. Sousa, "Fundamental limits of spectrum-sharing in fading environments," *IEEE Trans. Wirel. Commun.*, vol. 6, no. 2, pp. 649–657, Feb. 2007, doi: 10.1109/TWC.2007.05447.

[10] T. W. Ban, W. Choi, B. C. Jung, and D. K. Sung, "Multi-user diversity in a spectrum sharing system," *IEEE Trans. Wirel. Commun.*, vol. 8, no. 1, pp. 102–106, Jan. 2009, doi: 10.1109/T-WC.2009.080326.

[11] L. Cao and H. Zheng, "Distributed spectrum allocation via local bargaining," in *2005 Second Annual IEEE Communications Society Conference on Sensor and AdHoc Communications and Networks, SECON 2005*, 2005, vol. 2005, pp. 475–486, doi: 10.1109/SAHCN.2005.1557100.

[12] J. Huang, R. A. Berry, and M. L. Honig, "Auction-based spectrum sharing," in *Mobile Networks and Applications*, 2006, vol. 11, no. 3, pp. 405–408, doi: 10.1007/s11036-006-5192-y.

[13] R. Etkin, A. Parekh, and D. Tse, "Spectrum sharing for unlicensed bands," *IEEE J. Sel. Areas Commun.*, vol. 25, no. 3, pp. 517–528, Apr. 2007, doi: 10.1109/JSAC.2007.070402.

[14] Z. Ji and K. J. R. Liu, "Belief-assisted pricing for dynamic spectrum allocation in wireless networks with selfish users," in *2006 3rd Annual IEEE Communications Society on Sensor and Adhoc Communications and Networks, Secon 2006*, 2006, vol. 1, pp. 119–127, doi: 10.1109/SAHCN.2006.288416.

[15] Z. Ji and K. J. R. Liu, "Cognitive radios for dynamic spectrum access - Dynamic spectrum sharing: A game theoretical overview," *IEEE Communications Magazine*, vol. 45, no. 5. pp. 88–94, May-2007, doi: 10.1109/MCOM.2007.358854.

[16] D. Niyato and E. Hossain, "Competitive pricing for spectrum sharing in cognitive radio networks: Dynamic game, inefficiency of nash equilibrium, and collusion," *IEEE J. Sel. Areas Commun.*, vol. 26, no. 1, pp. 192–202, Jan. 2008, doi: 10.1109/JSAC.2008.080117.

[17] X. Kang, Y. C. Liang, H. K. Garg, and L. Zhang, "Sensing-based spectrum sharing in cognitive radio networks," *IEEE Trans. Veh. Technol.*, vol. 58, no. 8, pp. 4649–4654, 2009, doi:

[18] R. Zhang and Y. C. Liang, "Exploiting multi-antennas for opportunistic spectrum sharing in cognitive radio networks," *IEEE J. Sel. Top. Signal Process.*, vol. 2, no. 1, pp. 88–102, Feb. 2008, doi: 10.1109/JSTSP.2007.914894.

[19] H. Xie, B. Wang, F. Gao, and S. Jin, "A Full-Space Spectrum-Sharing Strategy for Massive MIMO Cognitive Radio Systems," *IEEE J. Sel. Areas Commun.*, vol. 34, no. 10, pp. 2537–2549, Oct. 2016, doi: 10.1109/JSAC.2016.2605238.

[20] Y. Zou, "Physical-Layer Security for Spectrum Sharing Systems," *IEEE Trans. Wirel. Commun.*, vol. 16, no. 2, pp. 1319–1329, Feb. 2017, doi: 10.1109/TWC.2016.2645200.

[21] R. Deng, J. Chen, C. Yuen, P. Cheng, and Y. Sun, "Energy-efficient cooperative spectrum sensing by optimal scheduling in sensor-aided cognitive radio networks," *IEEE Trans. Veh. Technol.*, vol. 61, no. 2, pp. 716–725, Feb. 2012, doi: 10.1109/TVT.2011.2179323.

[22] R. Xie, R. Yu, H. Ji, and Y. Li, "Energy-efficient resource allocation for heterogeneous cognitive radio networks with femtocells," *IEEE Trans. Wirel. Commun.*, vol. 11, no. 11, pp. 3910–3920, 2012, doi: 10.1109/TWC.2012.092112.111510.

[23] S. Bayhan and F. Alagöz, "Scheduling in centralized cognitive radio networks for energy efficiency," *IEEE Trans. Veh. Technol.*, vol. 62, no. 2, pp. 582–595, 2013, doi: 10.1109/TVT.2012.2225650.

[24] X. Huang, T. Han, and N. Ansari, "On green-energy-powered cognitive radio networks," *IEEE Commun. Surv. Tutorials*, vol. 17, no. 2, pp. 827–842, Apr. 2015, doi: 10.1109/COMST.2014.2387697.

[25] M. Costa and A. Ephremides, "Energy Efficiency Versus Performance in Cognitive Wireless Networks," *IEEE J. Sel. Areas Commun.*, vol. 34, no. 5, pp. 1336–1347, May 2016, doi: 10.1109/JSAC.2016.2520219.

[26] S. K. Leung-Yan-Cheong and M. E. Hellman, "The Gaussian Wire-Tap Channel," *IEEE Trans. Inf. Theory*, vol. 24, no. 4, pp. 451–456, 1978, doi: 10.1109/TIT.1978.1055917.

[27] J. Barros and M. R. D. Rodrigues, "Secrecy capacity of wireless channels," in *IEEE International Symposium on Information Theory - Proceedings*, 2006, pp. 356–360, doi: 10.1109/ISIT.2006.261613.

[28] N. Zhang, N. Lu, N. Cheng, J. W. Mark, and X. S. Shen, "Cooperative spectrum access towards secure information transfer for CRNs," *IEEE J. Sel. Areas Commun.*, vol. 31, no. 11, pp. 2453–2464, 2013, doi: 10.1109/JSAC.2013.131130.

[29] S. Goel and R. Negi, "Guaranteeing secrecy using artificial noise," *IEEE Trans. Wirel. Commun.*, vol. 7, no. 6, pp. 2180–2189, Jun. 2008, doi: 10.1109/TWC.2008.060848.

[30] E. Tekin and A. Yener, "The general Gaussian multiple-access and two-way wiretap channels: Achievable rates and cooperative jamming," *IEEE Trans. Inf. Theory*, vol. 54, no. 6, pp. 2735–2751, Jun. 2008, doi: 10.1109/TIT.2008.921680.

[31] L. Lai and H. El Gamal, "The relay-eavesdropper channel: Cooperation for secrecy," *IEEE Trans. Inf. Theory*, vol. 54, no. 9, pp. 4005–4019, 2008, doi: 10.1109/TIT.2008.928272.

[32] D. Xu and Q. Li, "Resource Allocation for Cognitive Radio With Primary User Secrecy Outage Constraint," *IEEE Syst. J.*, vol. 12, no. 1, pp. 839–904, Mar. 2018, doi: 10.1109/JSYST.2016.2585654.

[33] D. Xu and Q. Li, "Improving physical-layer security for primary users in cognitive radio networks," *IET Commun.*, vol. 11, no. 15, pp. 2303–2310, Oct. 2017, doi: 10.1049/iet-com.2017.0323.

[34] Q. Li and D. Xu, "Minimizing secrecy outage probability for primary users in cognitive radio networks," *AEU - Int. J. Electron. Commun.*, vol. 83, pp. 353–358, Jan. 2018, doi: 10.1016/j.aeue.2017.10.006.

[35] Y. C. Liang, Y. Zeng, E. C. Y. Peh, and A. T. Hoang, "Sensing-throughput tradeoff for cognitive radio networks," *IEEE Trans. Wirel. Commun.*, vol. 7, no. 4, pp. 1326–1337, Apr. 2008, doi: 10.1109/TWC.2008.060869.